\documentclass{aa}  
\usepackage{graphicx,amssymb}
\usepackage[varg]{txfonts}
\begin{document}
   \title{SALT long-slit spectroscopy of LBQS  2113-4538: variability of the Mg II and Fe II component\thanks{based on observations made with the Southern African Large Telescope (SALT) under program 2012-1-POL-008 (PI: B. Czerny)}\fnmsep\thanks{new Fe II template shown in Fig.~\ref{fig:templates} is only available in electronic form at the CDS via anonymous ftp to cdsarc.u-strasbg.fr (130.79.128.5) or via http://cdsweb.u-strasbg.fr/cgi-bin/qcat?J/A+A/}. }

   \author{K. Hryniewicz\inst{1} \and B. Czerny\inst{1} \and  W. Pych\inst{1} \and Udalski\inst{2} \and M. Krupa\inst{3} \and A. \' Swi\c eto\' n\inst{3} \and J. Kaluzny\inst{1} 
}

   \institute{Nicolaus Copernicus Astronomical Center, Bartycka 18, 00-716 Warsaw, Poland 
\and
Warsaw University Observatory, Al. Ujazdowskie 4, 00-478 Warszawa, Poland
\and
 Astronomical Observatory of the Jagiellonian University, Orla 171, 30-244 Cracow, Poland
             }

   \date{Received ............; accepted ..............}
  \abstract
   {The Mg II line is of extreme importance in intermediate redshift quasars since it allows us to measure the black hole mass in
these sources and to use these sources as probes of the distribution of dark 
  energy in the Universe, as a complementary tool to SN Ia.}
   {Reliable use of Mg II requires a good understanding of all the systematic effects involved in the measurement of the line
properties, including the contamination by Fe II UV emission.}
   {We performed three spectroscopic observations of a quasar LBQS 2113-4538 (z = 0.956) with the SALT telescope separated in time by
several months and we analyze in detail the mean spectrum and the variability in the spectral shape.}
   {We show that even in our good-quality spectra the Mg II doublet is  well fit by a single Lorentzian shape. We
   tested several models of the Fe II pseudo-continuum and showed that one of them well represents all the data. 
   The amplitudes of both components vary in time, but the shapes do not change significantly. The measured line width of 
   LBQS  2113-4538 identifies this object
   as a class A quasar. The upper limit of 3\% for the contribution of the Narrow Line Region (NLR) to Mg II may suggest that the 
  separation of 
   the Broad Line Region (BLR) and NLR disappears in this class of objects.}
   {}


\authorrunning{Hryniewicz et al.}
\titlerunning{SALT Long-slit Spectroscopy of LBQS  2113-4538}

   \maketitle
%

\section{Introduction}
Broad emission lines in active galactic nuclei (AGN) allow us to study the central parts {of the nuclei that cannot be spatially resolved with current instruments}. The lines probe the geometry, the inflow/outflow of the gas, and offer a way to measure the black hole mass relatively accurately. It was recently proposed that they can even be used to measure the expansion rate of the Universe since they obey a relation between the size of the emitting region and the absolute monochromatic luminosity (Watson et al. 2011, Czerny \& Hryniewicz 2011). The most important emission lines are H$\beta$ for nearby sources, Mg II for intermediate redshift objects, and CIV for high redshift quasars. Detailed studies of these lines are thus essential.

Extensive monitoring of over 50 nearby
quasars and Seyfert galaxies showed that H$\beta$ line intensity follows closely the variation of the continuum emission
with very little scatter, and the line width is mostly determined by the Keplerian motion without 
significant inflow/outflow pattern (e.g., Peterson 1993, Kaspi et al. 2000, Peterson et al. 2004, Wandel et al. 1999, Bentz et al. 2013). 
This line is now broadly used for black hole mass determination. However, 
for sources at higher redshifts the H$\beta$ line moves out of the optical band to the IR and thus has to be replaced
with another line.

In the intermediate redshift quasars the Mg II 2800 \AA  ~ line is an attractive option. The Mg II line, like H$\beta$, belongs to the low ionization line (LIL) class. These lines are thought to be emitted close to the accretion disk surface while high ionization lines (HIL), like CIV,  
come instead from outflowing wind and are strongly affected by varying conditions (Collin-Souffrin et al. 1988). There have been only only a few examples of Mg II monitoring in nearby objects (Clavel et al. 1991 for NGC 5548; Reichert et al. 1994 for NGC 3783; Peterson et al.
2004 for Fairall 9; Metzroth
et al. 2006 for NGC 4151) . The line was used instead of H$\beta$ to determine the black hole mass from a single epoch spectra in quasar samples by Kong et al. (2006) and  Vestergaard \& Osmer (2009).  Shen et al. (2008) analysis showed that Mg II is indeed a good proxy for the H$\beta$ line. On the other hand, Wang et al. (2009) 
on the basis of the study of SDSS intermediate redshift quasars concluded that the velocity width of Mg II tends to be smaller than that of 
H$\beta$. In their fits, the authors were using the multiple component fit to Mg II and the Fe II pseudo-continuum template from Tsuzuki et al. (2006).

In this paper we present high-quality medium resolution spectra of the quasar LBQS 2113-4538 ($z = 0.946 \pm 0.005$; Hewet et al. 1995) 
obtained with the Southern African
Large Telescope. The source was first identified in the objective-prism plates from the UK Schmidt Telescope (Morris et al. 1991)
and included in the Large Bright Quasar Sample (LBQS; Hewett et al. 1995) as well as in the Veron-Cetty \& Veron (2001) catalog. 
The previous measurements of the line equivalent width (EW) and the full width at half maximum (FWHM) indicated
typical values of the Mg II line width and shape,  and a simple line profile 
(EW = $27^{+14}_{-11}$ \AA, FWHM = $4500 \pm 1100$ km s$^{-1}$ in Forster et al. 2001, and $4500 \pm 550$  km s$^{-1}$ in 
Vestergaard \& Osmer 2009  for this source,
and in the whole LBQS sample the median values were EW = 33.8 \AA, FWHM = 4440  km s$^{-1}$ respectively, Forster et al. 2001). 
We analyze in detail the shape and the variability of the Mg II line and the
underlying Fe II emission in this source.

\section{Observations}
We performed three observations of the quasar LBQS 2113-4538  using the Robert Stobie Spectrograph 
(RSS; Burgh et al. 2003, Kobulnicky et al. 2003; Smith et al. 2006) on the Southern
African Large Telescope (SALT) in the service mode. The data was collected in three
blocks, on the nights of May 15/16, July 30/31, and November 18/19, 2012 (UT) thus covering a period of 6 months.
Each block
consisted of a pair of 978 second exposures of the target spectrum in a long slit mode, with
the slit width of 2", followed by an exposure
of the spectrum of the Argon calibration lamp, and a set of flat-field images. We used RSS PG1300 grating,
with the spectral resolution of $R=1047$ at 5500 \AA, and the PC04600 filter.
The observations were performed in a dark moon, bright moon, and grey moon conditions.
The nights were
not photometric, in particular the third (grey moon) observation was performed in the presence
of thin clouds.
Initial data reduction steps (gain correction, cross-talk correction, overscan bias subtraction, and amplifier mosaicking) 
were performed by the SALT Observatory staff
using a semi-automated pipeline from the SALT PyRAF package\footnote{http://pysalt.salt.ac.za} (see Crawford et al. 2010).
Flat-field correction and further reduction steps were performed using procedures
within the IRAF package. The pairs of adjacent spectrum images of LBQS 2113-4538
were combined into a single image. This enabled us to efficiently reject cosmic rays
and raise the signal-to-noise ratio.
Identification of the lines in the calibration lamp spectrum, wavelength calibration,
image rectification, and extraction of one dimensional spectra were done using 
functions from the noao.twodspec package within IRAF\footnote{IRAF is distributed by the National 
Optical Astronomy Observatories, which are operated
by the Association of Universities for Research in Astronomy, Inc., under cooperative agreement
with the NSF.}.

The spectroscopic observations were supplemented with the seven photometric observations in V band by the OGLE team (Fig.~\ref{fig:lightcurve}).
Broadband optical observations of LBQS 2113-4538 were carried out as a
subproject of the OGLE survey. The 1.3-m Warsaw telescope located at Las
Campanas Observatory, Chile, equipped with the 32 CCD detector mosaic
camera was used. The source LBQS 2113-4538 was observed approximately every two
weeks through the V-band filter with the exposure time of 240 seconds.
Collected images were reduced with the standard OGLE pipeline based on
the image difference technique (Udalski 2003). Accuracy of each single
measurement of LBQS 2113-4538 was about 0.01 mag. Additionally, the
stability of the photometry was checked by comparison of three nearby
brighter stars (V: 14.5-16.5 mag). All these stars were constant at
the 5 mmag level. 
The source was weakly but significantly variable, as measured with respect to the three comparison stars.

Here we analyze in detail the limited spectral band between the 2700  \AA~ and 2900 \AA~ (measured in the rest frame)
with the aim of reproducing the behavior of the Mg II and Fe II emission.

To perform flux calibrations we used the star TYC 8422-788-1 from the Hipparcos catalog located  in the slit and separated by 3.5 arc min from 
the quasar. The star, according to Piquard et al. (2001), is a variable star\footnote{http://cdsarc.u-strasbg.fr/cgi-bin/nph-Cat/html/max=107?II/233/tableb-v.dat} 
of the $\delta$ Cephei type, with a V-band luminosity of 11 mag and a period of 8.740332 days. The temperature of these stars typically varies between 5000 K and 6000 K, and the surface gravity is on order of 1.5 - 2.3.
We extracted the spectra of the star with the same procedure as for the quasar. Next we used the Castelli \& Kurucz (2004) model assuming an effective temperature 5500 K, surface gravity 1.5, and solar metalicity. We divided the observed spectrum of the star by the atmosphere model. The ratio was fitted with the third order polynomial thus creating the response function. 
The wavelength range above 5000 K was in the Wien tail and the models gave the same spectral 
shape, independent of assumptions of the star temperature and gravity.

This response function was then applied to each of the quasar spectra in the 5100 - 5700 \AA~ range in the observed frame. In the actual data fitting, only the narrower region (2700 - 2900 \AA~ in the rest frame;  5280 - 5672 \AA~ in the observed frame) was used, wide enough to cover the Mg II well but smaller than the whole region in order to minimize the systematic errors connected with the flux calibration.

We estimated the instrumental broadening close to the Mg II  
using the 5200 \AA ~ sky line to be $135 \pm 7$ km s$^{-1}$ for the lowest quality observation 3. This value 
is much narrower than the expected line width so in the data fitting we neglect this 
effect. We checked in one example that introducing the instrumental broadening leads
to marginally narrower lines by $\sim 100$  km s$^{-1}$.

Finally, the spectra were dereddened to account for the Galactic 
extinction assuming $A_{\lambda} = 0.150, 0.113$, and 0.090 in the B,V and R bands for this source
(Schlafly \& Finkbeiner 2011) 
after NED\footnote{NASA/IPAC Extragalactic Database (NED) is operated by the Jet Propulsion 
Laboratory, California Institute of Technology}, with quadratic extrapolation between 
these values for other wavelengths.

Next, we obtained the mean spectrum by adding the three spectra together. We then analyzed both the mean spectrum  
and the individual spectra in search of the variability in the spectral shape.

We neglected the intrinsic absorption as there is no clear signature of such an extinction
in the spectra. We also neglected a possible host galaxy contribution as it is not likely 
to be important at such short wavelengths.

\section{Model}
We modeled the spectrum assuming the following components: power-law continuum, Fe II pseudo-continuum, and
Mg II line. The line was modeled either as a single line at 2800 \AA~ or as a doublet (2796.35, 2803.53; Morton 1991). The doublet
ratio was found to be 1.2:1 in I Zw I FOS/HST data implying an optically thick case, and in our considerations we fixed this ratio
at 1:1.  The redshift of the object is treated as an arbitrary parameter since the determination 
by Forster et al. (2001) is not accurate from the point of view of our high-quality spectroscopy. 
The 2700 - 2900 \AA~ region is also contaminated by the Balmer
continuum (see, e.g., Dietrich et al. 2001) but this component is shallow in the limited wavelength range
so this component cannot be 
distinguished from the underlying power law. This effect slightly influences the fitted power-law slope
but does not affect either the normalization of the Fe II component or the Mg II shape.

A single Mg II component is modeled either as a Gaussian, or a Lorentzian. We also tried a two-component Guassian fit, as well as 
a rotationally broadened Lorentzian, following Kollatschny \& Zetzl (2013).

The Fe II UV emission is modeled with the use of several templates, both theoretical and observational. We start with 
the Vestergaard \& Wilkes (2001) template, based on I Zw 1 with zero contribution 
underlying the Mg II core, and the template of Tsuzuki et al. (2006) also based on I Zw 1 but with some Fe II
emission underlying Mg II obtained by subtraction of the two Gaussians modeling the Mg II doublet. Next we
experiment with purely theoretical templates of Bruhweiler and Verner (2008), calculated for different values of the density, 
turbulent velocity, and ionization parameter $\Phi$.  We allow for Fe II broadening as well as for the shift with respect to
 Mg II since Fe II was suggested to come from infalling material (Ferland et al. 2009).

\section{Analysis of the spectra}
\begin{figure}
   \centering
   \includegraphics[width=0.9\hsize]{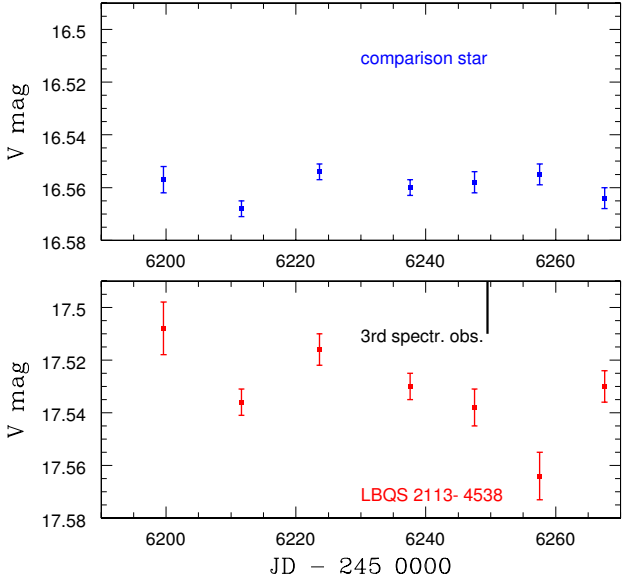}
   \caption{The V-band light curve of LBQS 2113-4538 (lower panel) and one of the comparison stars (upper panel); the last spectroscopic observation was performed during this photometric campaign.}
              \label{fig:lightcurve}%
    \end{figure}

We model the three high-quality RSS SALT spectra of LBQS  2113-4538 in the 2700 - 2900 \AA~ rest frame 
as consisting of the power-law continuum, the Fe II pseudo-continuum, and the Mg II line. The observation span covers six months. The photometric observations, covering three months, show clear although not very strong variations, with an amplitude of $\sim 0.03$ mag. The light curve and one of the three comparison stars are shown in  Fig. \ref{fig:lightcurve}.

%
%

\subsection{Mean spectrum}
Since the variability is not very strong, we started with the composite of the three spectra. The fits are summarized in Table~\ref{table:mean}. Line and pseudo-continuum EW was calculated in the 2700 - 2900 \AA~ band where the fit was performed, and both were calculated with respect to the power law continuum. The statistical errors for the best-fit parameters are typically on the order of 0.3 \AA~ for EW(Mg II), 2.0 \AA~ for EW(Fe II), and 50 - 100 km s$^{-1}$ for Mg II kinematic width. The Fe II smearing velocity was mostly fixed, but we tested a few possible values, and the implied accuracy of the values given in the table is about  100 - 200  km s$^{-1}$.

\begin{figure}
   \centering
   \includegraphics[width=0.95\hsize]{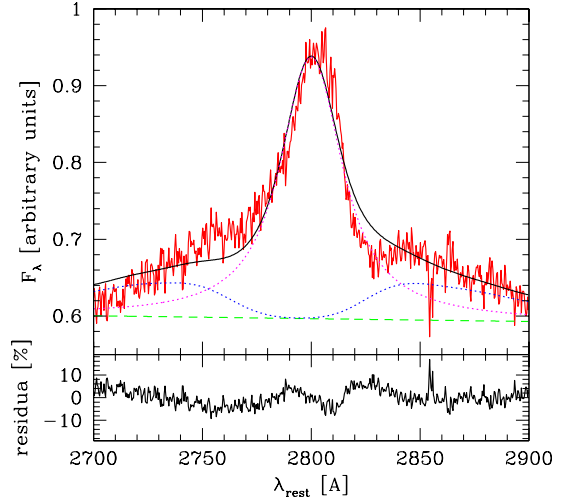}
   \caption{The best fit and the fit components for model C. Upper panel: Total model (black solid line),  continuum (green dashed line), continuum with Fe II pseudo-continuum (blue dotted line), Mg II component with continuum (red dotted line). Lower panel: residuals. See Table~\ref{table:mean} for the model parameters.}
              \label{fig:Cmean}%
    \end{figure}

The frequently used Fe II pseudo-continuum template of Vestergard \& Wilkes (2001), combined with Gaussian or Lorentzian shape of the Mg II line did not provide the satisfactory fit to the data. In Fig. \ref{fig:Cmean} we show the best fit for a Lorentzian shape, a doublet with the equal intensity of the components (model C), together with residuals. Strong residuals are clearly seen. The use of the templates from Tsuzuki et al. (2006) only partially reduces the residuals.

%

   \begin{figure}
   \centering
   \includegraphics[width=0.95\hsize]{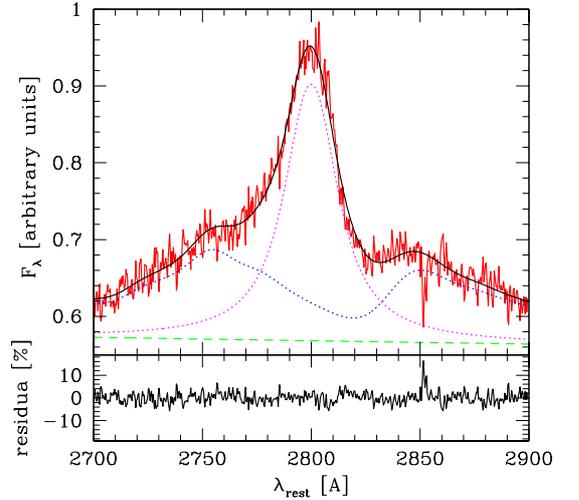}
   \caption{The best fit and residuals for the mean spectrum, as in Fig.~\ref{fig:Cmean} but for model E.}
              \label{fig:Emean}%
    \end{figure}
%
%

However, we obtained very interesting results for the theoretical templates of Fe II from  Bruhweiler and Verner (2008), which were recently applied to a large sample of SDSS quasars (Marziani et al. 2013). These templates give systematic residuals at 2950 \AA~ due to the absence of Fe I emission in the model (see Marziani et al. for the discussion), but our fits did not extend to such wavelengths.
The best fit for model E was fully satisfactory,  no significant patterns are seen (see Fig~\ref{fig:Emean}), and the $\chi^2/dof$ dropped by almost a factor of three. It has been frequently argued that the Lorentzian shape provides a very good description and our high-quality spectra support this conclusion. All the complexities of the nearby wavelength range are satisfactorily explained by the complexity of the Fe II component. In particular, the strong pattern at $\sim 2750$ \AA ~ and the somewhat weaker feature at $\sim 2850 $ \AA~ in the total spectrum are then nicely reproduced. The exact parameters of the best theoretical template (model E) are: model with 735 atomic levels, number density $n_H = 10^{11}$ cm$^{-3}$, turbulent velocity $v_{turb} = 20$ km s$^{-1}$, and $\Phi = 10^{20.5}$ cm$^{-2}$ s$^{-1}$. Interestingly, Bruhweiler and Verner (2008) came to the same conclusion when analysing the I Zw I spectrum.

The fits for several other theoretical templates from Bruhweiler and Verner (2008) were not as good, but were still reasonable (models F and L). The worst fit was for  models J and K which suggests that  $v_{turb} > 10$ km s$^{-1}$. The comparison of model L and P strongly favors densities lower than $10^{12}$ cm$^{-3}$.

We tested whether allowing for a velocity shift of the Fe II pseudo-continuum and the Mg II line improves the fit. However, we have not found any evidence of such a difference, we only obtained the upper limit of 300 km s$^{-1}$ for the relative velocity.

We also checked whether the rotationally broadened Lorentzian shape, justified theoretically (see, e.g., Kollatschny \& Zetzl 2013) provides a better fit than a single Lorentzian.  However, when we considered the intrinsic width of the Lorentzian to be much narrower than the rotational broadening, the fit quality dropped. Better fits than for a single Lorentzian were obtained when the rotational broadening was smaller than the intrinsic width, but the change was not actually significant (the drop in $\chi^2$ below 2.0). This does not mean that there is no rotational component in the line; it only shows that this rotational broadening is not well reproduced by the orbital motion of the spherically distributed clouds.

\begin{table*}
\caption{Parameters of the fits for the mean spectrum. }   
\label{table:mean}      
\centering                          
\begin{tabular}{l r r r r r r r r r r}        
\hline\hline      
Model       &  S/D &  Line shape & Fe II template & Fe II & slope & z & Mg II & Mg II & Fe II&$\chi^2$  \\
            &      &             &                & Width &       &   & EW    & FWHM &EW               \\ 
            &      &             &                &km s$^{-1}$  &       &   & \AA        & km s$^{-1}$& \AA  &   \\  
\hline                        
A           &  Single         &  Gauss      &  Vestergaard  & 1200 & -0.13  &0.954& 14.64  &  3530       &1.47 &1680.9    \\   
B           &  Single         &  Lorenzian  &  Vestergaard  & 1200 & -0.15  &0.954&  20.69 &  3200       & 0.18 &1230.0    \\ 
C           &  Doublet        &  Lorenzian  &  Vestergaard  & 1200 & -0.14  &0.954&  20.80 &  3140       & 0.27 &1238.5    \\ 
D           &  Single         &  Lorenzian  &  d11-m20-20.5-735 & 900 & -0.22 & 0.956 & 24.78     &  3050       & 25.62  &520.2  \\
E           &  Doublet        &  Lorenzian  &  d11-m20-20.5-735 & 900 & -0.22 & 0.956 &   23.81      &  2800       & 24.77   &507.3    \\
F           &  Doublet        &  Lorenzian  &  d11-m30-20-5-735 & 1100 & -0.20 & 0.956 & 23.61      &  2800       & 23.89  &530.0  \\ 
G           &  Doublet        &  Lorenzian  &  Tsuzuki          & 900 &  -0.30 & 0.956 & 22.85   &   2600 &   23.46 & 891.9 \\
H           &  Doublet        &  Lorenzian  &  d11-m20-21-735     & 1100 & -0.23 & 0.956 & 23.53      &  2750       & 25.04  &537.2 \\
I           &  Doublet        &  Lorenzian  &  d10-5-m20-20-5     & 1100 & -0.02 & 0.956 & 23.02      &  2800       & 23.11  &582.5 \\
J           &  Doublet        &  Lorenzian  &  d11-m05-20-5     & 1100 & 0.00     & 0.956 & 22.34      &  2800       & 18.86  &608.4 \\
K           &  Doublet        &  Lorenzian  &  d11-m10-20-5     & 1100 & 0.13    & 0.956 & 22.85      &  2800       & 21.87  &594.4 \\
L           &  Doublet        &  Lorenzian  &  d11-m20-20-5     & 1100 & 0.25    & 0.956 & 23.52      &  2800       & 26.33  &533.1 \\
M           &  Doublet        &  Lorenzian  &  d11-m30-20-5     & 1100 & 0.32    & 0.956 & 24.56      &  2900       & 28.34  &552.7 \\
N           &  Doublet        &  Lorenzian  &  d11-m50-20-5     & 1100 & 0.47    & 0.956 & 25.52      &  2900       & 32.88  &541.4 \\
O           &  Doublet        &  Lorenzian  &  d11-5-m20-20-5     & 1100 & 0.20  & 0.956 & 23.97      &  2850       & 24.07  &570.6 \\
P           &  Doublet        &  Lorenzian  &  d12-m20-20-5     & 1100 & 0.24    & 0.956 & 23.89      &  2800       & 22.11  &567.6 \\
Q           &  Doublet        &  Lorenzian  &  d11-m20-20     & 1100   & 0.23  & 0.956 & 23.67      &  2800       & 24.29  &560.3 \\
R           &  Doublet        &  Lorenzian  &  d11-m20-21     & 1300   & 0.31    & 0.956 & 23.93      &  2800       & 29.77 &  559.4 \\
\hline                                   
\end{tabular}
\end{table*}

\subsection{Separate spectra}
We next analyze the three spectra separately because the observational conditions were significantly different in all three observations, and the source did vary, as shown by the photometric measurements.

\begin{figure}
   \centering
   \includegraphics[width=0.95\hsize]{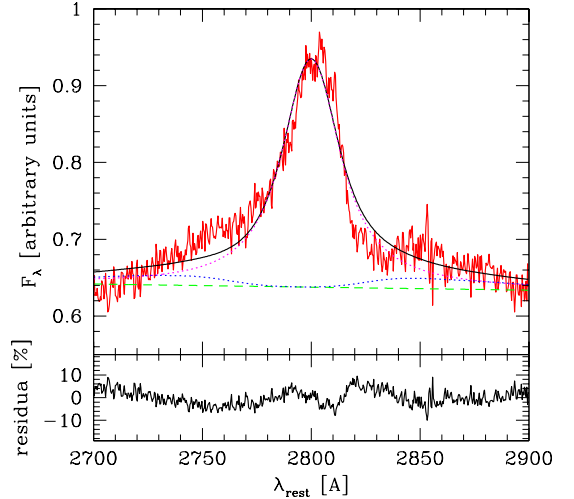}
   \caption{The best fit and residuals for observation 1, model C (see Table~\ref{table:three} for model parameters and Fig.~\ref{fig:Cmean} for the description of individual curves).}
              \label{fig:Cpierwsze}%
    \end{figure}

\begin{figure}
   \centering
   \includegraphics[width=0.95\hsize]{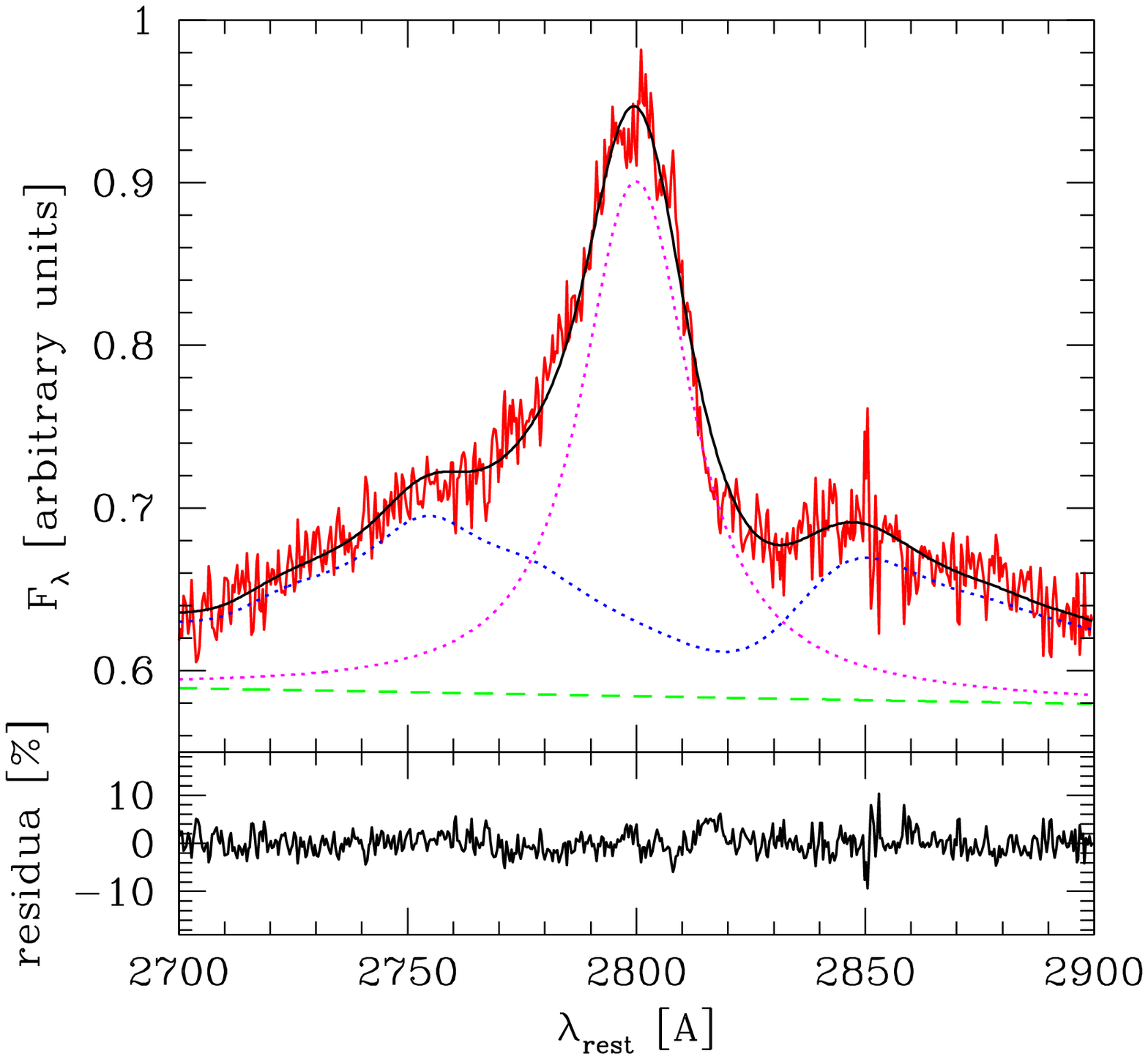}
   \caption{The best fit and residuals for observation 1, model E (see Table~\ref{table:three} for model parameters and Fig.~\ref{fig:Cmean} for the description of individual curves).}
              \label{fig:Epierwsze}%
    \end{figure}

\begin{figure}
   \centering
   \includegraphics[width=0.95\hsize]{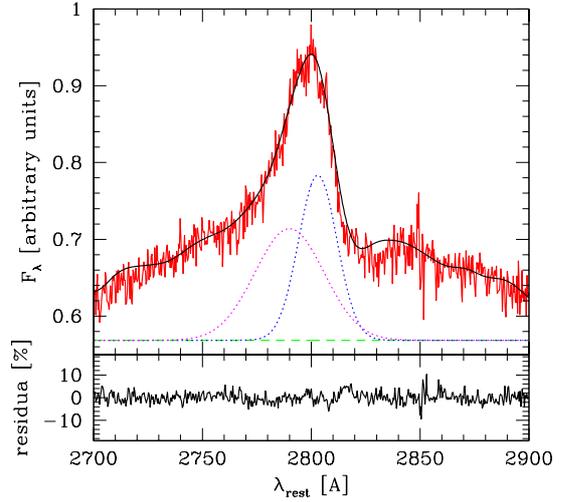}
   \caption{The best fit and residuals for the Tsuzuki et al. Fe II template and the two-Gaussian model for Mg II LBQS 2113-4538 in observation 1.}
              \label{fig:tsuzuki}%
    \end{figure}

The first (i.e., earliest) spectrum is of the highest quality (signal to noise (S/N) ratio $\sim 50$), obtained in the dark moon condition, with the sky background a factor of almost 3 lower than the continuum flux. 
We used the same models as before. Again, the Fe II template of Vestergaard \& Wilkes (2001) gave a poor fit and clearly visible residua (see Fig.~\ref{fig:Cpierwsze}) while the use of theoretical templates reduced all residua very strongly. The best fit was obtained for model E, as in the case of the mean spectrum. In Fig.~\ref{fig:Epierwsze}  we show the best fit and the residua, with statistical errors giving $\chi^2/dof =   684.4/562$. Systematic errors are too difficult to estimate, particularly in the case of a complex instrument like SALT. However, the residuals do not show any pattern. Large deviations at 2850 \AA~(rest frame) are due to imperfect subtraction of the sky line. Slightly worse fits are again obtained for models F and L, and the worst fits again for models J and K. Better fits for other templates can be obtained assuming at least two separate components for the Mg II line. For example, in Fig.~\ref{fig:tsuzuki} we show such a two-component Mg II fit combined with the Tsuzuki et al. (2006) template. The fit quality is only slightly worse than for a single-component model E ($\chi^2$ of 712.9 vs. 684.4). The two Mg II components have  FWHM of 1600 and 2890 km s$^{-1}$, and they are blueshifted in velocity space by 1390 km s$^{-1}$. Both components are then comparably broad, so the satisfactory interpretation of this fit is rather difficult, since none of the components correspond to the contribution from the NLR. The statistical errors for the best-fit parameters in these data sets are typically on the order of 0.1 \AA~ for EW(Mg II), lower than for the mean spectrum.

We also checked in this case whether the conclusion about the Lorentzian shape of the spectrum depends on the template. With this purpose, we also combined the best template, as in model E, with the Gaussian line for Mg II, assuming either a single Gauss, or a doublet. In both cases (models S and T in Table~\ref{table:three}) the fit was much worse than for model E.

%
%

%
%

%
%

The second spectrum, obtained under the bright moon conditions, has much higher sky background flux, dominating the source by a factor of 8 (S/N $\sim 17$). Statistical errors are therefore the dominant source of the error, and satisfactory fits can be obtained.
The Fe II template  of Vestergaard \& Wilkes (2001)  did not represent the data well,  but theoretical models again gave interesting results. The best fit was for model F., i.e., for the case of somewhat higher turbulent velocity  ($v_{turb} \sim 30$ km s$^{-1}$ instead of 20) than in the first observation. However, the $\chi^2$ for model E was only lower by 4.1, so within a two-sigma error. The worst solutions were again for models J and K. The line kinematic width measured by various models in observation 2 was slightly higher than in observation 1, but the change was within two-sigma error. The EW of the line clearly increased, in most models rising by $\sim 25$ \%, and the EW of the Fe II component also increased. Detailed assesment of the change in these two components is postponed to Sect.~\ref{sect:EW_var}.  The statistical errors for the best-fit parameters in observation 2 are typically on the order of 0.3 \AA~ for EW(Mg II), comparable to the results for the mean spectrum.

The third spectrum was obtained in the most difficult conditions, the grey moon with some intervening clouds. The background did not dominate the source as much as in the previous case, only by a factor of 3,  nevertheless, the S/N ratio in this data set is $\sim 14$. 
Statistical errors are very large; the EW(Mg II) is determined with a large error (for example, for model E, we have $23.79^{+2.26}_{-2.07}$ \AA). Systematic errors are also large since the signal directly measured by the telescope was a factor of 3 lower because of the cloud extinction. However, model E was the best again, and model N the worst of the models based on theoretical templates. The line was relatively weak during this observation, and the spectrum was much steeper.

None of the spectra shows any narrow intrinsic absorption features.

\begin{table*}
\caption{Parameters of the fits for the three individual spectra obtained with SALT between May and November 2012.}   
\label{table:three}      
\centering                          
\begin{tabular}{l r r r r r r r r r r}        
\hline\hline      
Model       &  S/D &  Line shape & Fe II template & Fe smear & slope & z & Mg II & Mg II & EW Fe II &$\chi^2$  \\
            &      &             &                &          &       &   &  EW   & FWHM  & EW  \\
            &                 &             &                &          &       &   & \AA        & km s$^{-1}$  &    \AA     &        \\  
\hline                        
Obs. 1 \\
\hline
A           &  Single         &  Gauss      &  Vestergaard  & 1200 & -0.11   & 0.954 &  14.23     &  3530     & 2.86   &  2167.4    \\ 
B           &  Single         &  Lorenzian  &  Vestergaard  & 1200 & -0.18   & 0.954 &  20.83   &  3350     & 2.32 &  1556.1        \\
C           &  Doublet        &  Lorenzian  &  Vestergaard  & 1200 & -0.18  & 0.954 &  20.54   &  3150     & 2.60 &  1569.9        \\
D           &  Single         &  Lorenzian  &  d11-m20-20.5-735 & 900 & -0.24& 0.956 &  21.78   &  2950     & 22.04 &  707.8     \\
E           &  Doublet        &  Lorenzian  &  d11-m20-20.5-735 & 900 & -0.24& 0.956 &  21.33   &  2690     & 22.48 &  684.4     \\
F           &  Doublet        &  Lorenzian  &  d11-m30-20-5-735 & 1100 & -0.16& 0.956 &  22.43  &  2750    & 26.88 &  712.9    \\
G           &  Doublet        &  Lorenzian  &  Tsuzuki    & 900  & -0.26  & 0.956 &  21.69  &  2500    & 26.74 &  1274.2   \\
H           &  Doublet        &  Lorenzian  &  d11-m20-21-735 & 1100 & -0.23  & 0.956 &  21.46  &  2680     & 23.66 &  741.2    \\
I           &  Doublet        &  Lorenzian  &  d10-5-m20-20-5 & 1100 & 0.28   & 0.956 &  22.31  &  2800     & 26.09 &  778.9    \\
J           &  Doublet        &  Lorenzian  &  d11-m05-20-5   & 1100 & 0.06  & 0.956 &  19.36  &  2650     & 15.39 &  863.4    \\
K           &  Doublet        &  Lorenzian  &  d11-m10-20-5   & 1100 & 0.18   & 0.956 &  21.29  &  2750     & 22.20 &  832.1    \\
L           &  Doublet        &  Lorenzian  &  d11-m20-20-5   & 1100 & 0.25   & 0.956 &  21.71  &  2750     & 26.70 &  728.5    \\
M           &  Doublet        &  Lorenzian  &  d11-m30-20-5   & 1100 & 0.25   & 0.956 &  22.15  &  2800     & 26.28 &  758.1    \\
N           &  Doublet        &  Lorenzian  &  d11-m50-20-5   & 1100 & 0.40   & 0.956 &  23.33  &  2900     & 30.22 &  741.4    \\
O           &  Doublet        &  Lorenzian  &  d11-5-m20-20-5 & 1100 & 0.16   & 0.956 &  21.79  &  2800     & 21.94 &  786.7    \\
P           &  Doublet        &  Lorenzian  &  d12-m20-20-5   & 1100 & 0.22  & 0.956 &  22.36  &  2800     & 21.70 &  778.1    \\
Q           &  Doublet        &  Lorenzian  &  d11-m20-20     & 1100 & 0.20  & 0.956 &  21.70  &  2750      & 23.28 &  756.3    \\
R           &  Doublet        &  Lorenzian  &  d11-m20-21     & 1300 & 0.30  & 0.956 &  21.72  &  2700    & 28.72 &  778.2    \\
S           &  Doublet        &  Gauss      &  d11-m20-20.5-735 & 900 & -0.19 & 0.956 & 14.43  &  2700    & 23.72 &  915.4  \\
T           &  Single         &  Gauss      &  d11-m20-20.5-735 & 900 & -0.19 & 0.956 & 14.40  & 2820     & 23.71 &  916.4  \\
\hline                                   
\hline                        
Obs. 2 \\
\hline
A           &  Single         &  Gauss      &  Vestergaard      & 1200 & 0.08  & 0.954 &  20.26  &  3940       & 9.55 &  777.5   \\
B           &  Single         &  Lorenzian  &  Vestergaard      & 1200 & 0.08  & 0.954 &  29.20  &  3750       & 7.61 &  649.5   \\
C           &  Doublet        &  Lorenzian  &  Vestergaard      & 1200 & 0.08   & 0.954 &  28.65  &  3500       & 8.11 & 653.8   \\
D           &  Single         &  Lorenzian  &  d11-m20-20.5-735 &  900 & -0.05  & 0.956 &  25.22  &  3050       & 21.03  &  484.2   \\
E           &  Doublet        &  Lorenzian  &  d11-m20-20.5-735 &  900 & 0.00 & 0.956 &  26.54  &  2870       & 26.65 &  476.3   \\
F           &  Doublet        &  Lorenzian  &  d11-m30-20-5-735 & 1100 & 0.05   & 0.956 &  26.05  &  2850       & 26.42 &  478.2   \\
G           &  Doublet        &  Lorenzian  &  Tsuzuki      &  900 & 0.03   & 0.956  &   28.16  &  2650 & 38.21 & 550.6 \\
H           &  Doublet        &  Lorenzian  &  d11-m20-21-735   & 1100 & 0.00   & 0.956 &  26.05  &  2850       & 26.34 &  478.8   \\
I           &  Doublet        &  Lorenzian  &  d10-5-m20-20-5   & 1100 & 0.50   & 0.956 &  26.01  &  2900       & 26.42 &  484.1   \\
J           &  Doublet        &  Lorenzian  &  d11-m05-20-5     & 1100 & 0.30   & 0.956 &  23.10  &  2750       & 17.05 &  497.2  \\
K           &  Doublet        &  Lorenzian  &  d11-m10-20-5     & 1100 & 0.44 & 0.956 &  25.29  &  2900       & 24.17 &  489.3  \\
L           &  Doublet        &  Lorenzian  &  d11-m20-20-5     & 1100 & 0.61   & 0.956 &  27.15  &  2900       & 31.92 &  478.7 \\
M           &  Doublet        &  Lorenzian  &  d11-m30-20-5     & 1100 & 0.61  & 0.956 &  27.72  &  3000       & 31.36 &  481.8 \\
N           &  Doublet        &  Lorenzian  &  d11-m50-20-5     & 1100 & 0.83  & 0.956 &  29.16  &  3050       & 37.15 &  480.5 \\
O           &  Doublet        &  Lorenzian  &  d11-5-m20-20-5   & 1100 & 0.38   & 0.956 &  25.88  &  2900       & 23.36 &  480.5 \\
P           &  Doublet        &  Lorenzian  &  d12-m20-20-5     & 1100 & 0.46  & 0.956 &  26.34  &  2900       & 22.78 &  485.1 \\
Q           &  Doublet        &  Lorenzian  &  d11-m20-20       & 1100 & 0.48   & 0.956 &  26.42  &  2920       & 25.99 &  484.7 \\
R           &  Doublet        &  Lorenzian  &  d11-m20-21       & 1300 & 0.61   & 0.956 &  27.08  &  2900       & 33.48 &  480.9 \\
\hline                                   
\hline                        
Obs. 3 \\
\hline
A           &  Single         &  Gauss      &  Vestergaard      & 1200 & -1.03 & 0.954 &  20.91  &  3825    & 14.00 &  373.0 \\
B           &  Single         &  Lorenzian  &  Vestergaard      & 1200 &  -1.11 & 0.954 &  31.74 & 3700    & 14.20 &  323.1 \\
C           &  Doublet        &  Lorenzian  &  Vestergaard      & 1200 &  -1.10 & 0.954 & 31.84 & 3600   & 14.05 &   325.0 \\
D           &  Single         &  Lorenzian  &  d11-m20-20.5-735 &  900 & -1.12 & 0.956 &  24.69  & 2950  & 21.49 & 237.7  \\
E           &  Doublet        &  Lorenzian  &  d11-m20-20.5-735 &  900 & -1.09 & 0.956  & 23.79 & 2650 & 22.43 & 237.3 \\
F           &  Doublet        &  Lorenzian  &  d11-m30-20-5-735 & 1100 & -1.09   & 0.956  &  24.22  &  2720 & 21.99 & 240.3 \\
G           &  Doublet        &  Lorenzian  &  Tsuzuki          &  900 & -1.13 & 0.956 &  24.550  &  2450 & 29.054 & 262.8 \\        
H           &  Doublet        &  Lorenzian  &  d11-m20-21-735   & 1100 & -1.07  & 0.956  &   28.48  &  2700 & 42.39 & 292.0 \\
I           &  Doublet        &  Lorenzian  &  d10-5-m20-20-5    & 1100 & -0.68  & 0.956  &  23.52   &  2650  & 22.74 & 242.9 \\
J           &  Doublet        &  Lorenzian  &  d11-m05-20-5     & 1100 & -0.80   & 0.956  &  21.63  & 2625  & 13.78 & 247.2 \\
K           &  Doublet        &  Lorenzian  &  d11-m10-20-5     & 1100 & -0.70 & 0.956 &  23.52  &  2650  & 22.40 &  248.7 \\
L           &  Doublet        &  Lorenzian  &  d11-m20-20-5     & 1100 & -0.45 & 0.956 & 24.74  &  2600  & 33.85 & 250.3 \\
M           &  Doublet        &  Lorenzian  &  d11-m30-20-5     & 1100 & -0.53  & 0.956 &  26.79 &  2900 &31.53& 248.5 \\
N           &  Doublet        &  Lorenzian  &  d11-m50-20-5     & 1100 &  0.35  &  0.956 & 36.12 & 3100 & 66.83 & 312.6 \\
O           &  Doublet        &  Lorenzian  &  d11-5-m20-20-5   & 1100 & -0.35 & 0.956 & 30.50 & 3000 & 40.41 & 295.9 \\
P           &  Doublet        &  Lorenzian  &  d12-m20-20-5       & 1100 & -0.45    & 0.956 & 27.33  & 2800 & 30.23 & 266.1 \\
Q           &  Doublet        &  Lorenzian  &  d11-m20-20         & 1100 & -0.70   & 0.956  & 24.30 & 2775 & 21.83 & 243.8 \\
R           &  Doublet        &  Lorenzian  &  d11-m20-21       & 1300 &  -0.70 & 0.956 & 23.48  & 2700 & 23.03 & 243.2 \\
\hline                                   
\end{tabular}
\end{table*}

\section{Results}
The three long-slit observations of the quasar done with the SALT telescope between May and November 2012 clearly showed the variability of the relative normalization of the three spectral components: the Mg II line, the Fe II pseudo-continuum, and the power-law continuum.

\subsection{Variations in EW of Mg II and Fe II and relative shifts}
\label{sect:EW_var}
We found that the photometric variations in the source are accompanied by the spectral variability. Selecting model E for a comparison of the three spectra, we
observed an increase in the EW(Mg II) of 24.4 \% between the first and the second observation, followed by a decline in the third; EW(Fe II) varied in a similar way. The ratio of EW(Mg II)/EW(Fe II) was constant within the measurement error, but the accuracy of the Fe II measurement is not very high (the ratio of EW(Fe II) in observation 2 to observation 1 is $1.19 \pm 0.1$).  The errors in observation 3 are still larger, so we cannot treat the conclusion about the common temporal variability of Mg II and Fe II with high confidence. The kinematic width of the Mg II line did not show any significant variations, with an upper limit of 14 \%. A similar lack of Mg II line (broad component) shape variability despite the variations of the line intensity was observed for NGC 3516 (Goad et al. 1999). Woo (2008) in a two-year campaign observed the change in the line kinematic width by 8 to 17 \%. The Mg II line is broader than the requested broadening of the Fe II template (1350 km s$^{-1}$ vs. 900 km s$^{-1}$) so Fe II may come from a somewhat larger distance. The slope of the underlying continuum also changed, but the change was not correlated with EW(Mg II).

We also looked for solutions allowing for an arbitrary kinematic shift between the position of Mg II and Fe II which would be interpreted as a signature of inflow or outflow for one of the components. In observation 1, the best fit was actually obtained for a relative velocity of $45^{+24}_{-33}$ km s$^{-1}$ (two-sigma error), indicating marginal hints of Mg II inflow if Fe II is in rest, or Fe II outflow if Mg II provides the reference. In the other two observations the data quality is too low to see any significant shift between Mg II and Fe II.

\subsection{Non-parametric variability test}
\begin{figure}
   \centering
   \includegraphics[width=0.95\hsize]{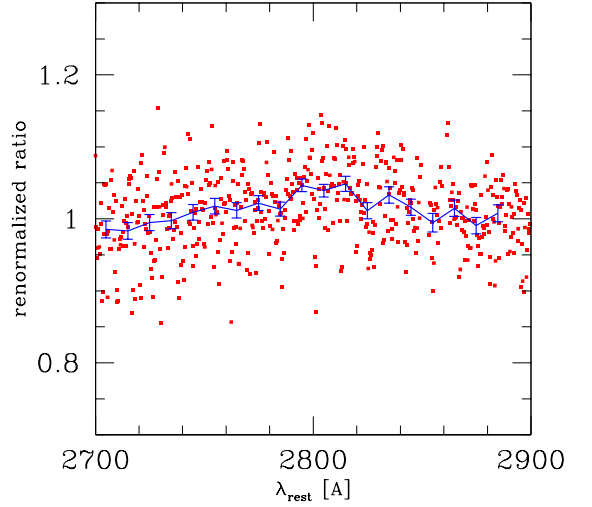}
   \caption{The normalized ratio of the data points from observation 2 to observation 1 (red points - original data, blue line with errorbars - data binned in wavelength by a factor of 30. We removed the points around 2850 \AA~ affected by the imperfect subtraction of the sky emission.}
              \label{fig:ratio}%
    \end{figure}

To confirm the overall spectral variability beyond any doubt, we tested the variations of the spectral shape of the quasar simply by dividing the second spectrum by the first one. The renormalized ratio, with the average value set to 1, shows considerable curvature due to the variations in Mg II and Fe II strongly dominating the spectrum in its middle part (see Fig.~\ref{fig:ratio}).  The $\chi^2/dof$ for a linear fit to the binned data points is 2.9, thus confirming the visual impression of the strong curvature due to the different contribution of Mg II to the two spectra. The slope of the linear fit is mostly sensitive to the intrinsic change of the slope of the continuum, since the Mg II and Fe II contributions are almost symmetric around 2800 \AA. This slope is positive, $(1.08 \pm 0.48) \times 10^{-4} $ \AA$^{-1}$, and if interpreted as a change solely due to the change in the slope, it implies the slope change of $0.3 \pm 0.14 $ between the two data sets. We stress, however, that the absolute value of the slope is certainly not well measured from our data since the exact value of the slope strongly depends on the Fe II template used in data analysis.

%
%

\subsection{Fe II observational template}

   \begin{figure}
   \centering
   \includegraphics[width=0.9\hsize]{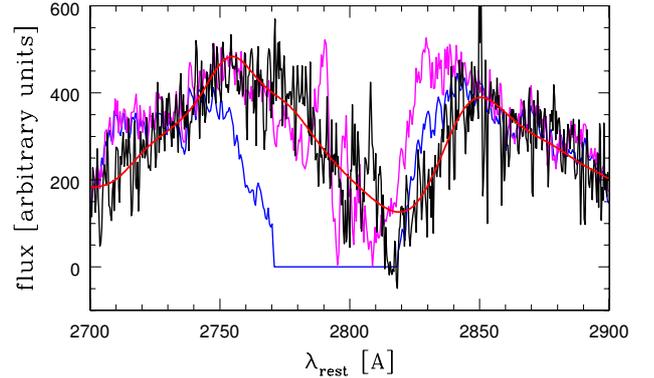}
   \caption{The comparison of various templates used to model the Fe II pseudo-contunuum in the UV band: Vestergaard \& Wilkes (2001) - blue line, 
Tsuzuki et al. (2006) - magenta line, our newly proposed template - black line, 11-m20-20.5-735 of Bruhweiler \& Verner (2008) convolved with 900 km/s Gaussian broadening, as used in the current paper - red thick line.}
              \label{fig:templates}%
    \end{figure}
%
%

The question of the Fe II contribution to the optical and UV spectrum has been recognized for a long time (Netzer \& Wills 1983; Collin-Souffrin \& Dumont 1986).
There has been noticeable effort to deal with this, but the question has not been resolved yet.

In the three observations very similar theoretical Fe II templates provided the best fit: it was d11-m20-20.5-735 for the first two observations, and a very similar template d11-m30-20-5-735 for the third, with the fit by d11-m20-20.5-735 giving $\chi^2$ worse only by 3.0. It is thus natural to assume that a universal template describes the data well. Theoretical templates are not yet perfect, as Fe II is an extremely complex ion, so observational templates have clear advantages.

We use the observation 1 data to determine a new Fe II template in the 2700 - 2900 \AA ~ range by subtracting the fitted power law and the Mg II 
doublet from the total spectrum. 
In our basic modeling we used the doublet ratio of 1:1 for Mg II, for simplicity, since the line is directly unresolved. 
However, this assumption could affect the derived template shape. For that purpose, we obtained  fits for several values of the Mg II
doublet ratio between 1:1 and 1.25:1, for the same theoretical Fe II template and broadening as above. The best fit was for the ratio of  1.15:1, and the 
total $\chi^2$ was reduced by 5.9, i.e., implying the difference of more than 2 standard deviations, in comparison with the 1:1 ratio. The largest differences
were in the red wing of Mg II where the deviations connected with the change in the doublet ratio from 1:1 to 1.15 were up to 15 \%. Therefore, we
considered the 1.15:1 ratio as the most appropriate to derive the template.
The parameters of the Mg II line in the total fit did not change within the error.

The comparison of this template with the previously used observational template and with the theoretical 
template 11-m20-20.5-735 of Bruhweiler \& Verner (2008) broadened with a 900 km s$^{-1}$ Gaussian are shown in Fig.~\ref{fig:templates}.
This template can be applied to sources with intrinsic broadening greater than $\sim 900 $ km s$^{-1}$. We removed four observational 
points close to 2850 \AA~ since in this region the extraction of the very strong sky line was not perfect, but this does not change the overall densly covered shape.

The main difference between the newly proposed template and the Tsuzuki et al. (2006) template is in the 2820 - 2840 \AA~ range. This would affect 
the modeling of the blue wing of the Mg II line. Otherwise, their template, obtained for a nearby object (I Zw 1), has a somewhat better S/N ratio.

The template may, in principle, contain the contribution from the MgII narrow component as it was not subtracted. However, as we discuss in more detail in Sect. 6.1, we do not detect this component in LBQS 2113-4538, despite the good quality of our data.

\subsection{Global parameters}

The determination of the Mg II line width opens a possibility to estimate the global parameters of the LBQS 2113-4538. 
We can estimate the black hole mass and the Eddington ratio is this object using various formulae applicable to the UV spectral range.
The SALT spectroscopic observation does not provide an absolute calibration. We thus adopt V mag of 17.29 from Hewett et al. (1995)
and correct it for the Galactic extinction $A_V = 0.113$ from NED. Using the standard cosmology ($H_o = 71$ km s$^{-1}$ Mpc$^{-1}$, 
$\Omega_m = 0.270$, $\Omega_{\Lambda} = 0.730$) we obtain a monochromatic luminosity of  $1.24 \times 10^{46}$ erg s$^{-1}$ at 2800 \AA.
We then apply the formulae based on the 3000 \AA ~ continuum and Mg II line measurements. The bolometric luminosity is 
$7.32 \times 10^{46}$ erg s$^{-1}$ for a bolometric correction factor of 5.9 (McLure \& Dunlop 2004). This value is very similar to the value of $8.13 \times 10^{46}$  erg s$^{-1}$ given by Vestergaard \& Osmer (2009). The FWHM of the Mg II line for our model E fits
in observation 1 is 2690 km s$^{-1}$. Using the formulae of Kollmeier et al. (2006),
McGill et al. (2007), or Trakhtenbrot \& Netzer (2012), we obtain the values for the black hole mass of $1.03 \times 10^9 M_{\odot}$, 
$4.72 \times 10^8  M_{\odot}$, and $8.05 \times 10^8  M_{\odot}$,  and the 
Eddington ratios of  0.57, 1.23, and  0.72, respectively. This ratio is somewhat higher than the value of 0.42 derived by Vestergaard \& Osmer (2009), mostly due to much narrower line kinematic width in our fits, so the corresponding black hole mass in Vestergaard \& Osmer (2009), $1.51 \times 10^9 M_{\odot}$, is higher than any of the values quoted above.

The estimates made above rely on the assumption that the motion of the BLR clouds is predominantly Keplerian. There were suggestions that the radiation pressure may be strong enough to modify the cloud dynamics, particularly for high Eddington ratio sources (e.g.,  Marconi et al. 2008).  This may induce a systematic error leading to smaller values of the black hole. It is difficult to make a quantitative statement about this question, but on the other hand the symmetry of the Mg II line in the observed source does not suggest a significant outflow velocity. This, however, is most likely a problem for projects of black hole mass determination based on C IV (see, e.g., Czerny et al. 2013 and the references therein).


\section{Discussion}
We analyzed three long-slit spectroscopic observations of the quasar LBQS 2113-4538 to determine the shape and the variability of the Mg II broad 
emission line and the Fe II pseudo-continuum.

The object shows clear variations in all spectral components. The Mg II line is measured very accurately in the first two observations, with an EW error of 
0.1 \AA~ and 0.3 \AA, respectively; in the third observation the measurement is of lower quality because of the clouds. The line varied by 25 \% during the six-month period. This shows 
that reverberation study of distant quasars can be done with the SALT telescope. The level of the Fe II pseudo-continuum is measured much less accurately since it
is strongly coupled to the varying slope and normalization of underlying continuum. The ratio of the Fe II and Mg II did not change significantly, but longer observations would be needed to give a firm statement on this question. The kinematic broadening is slightly higher for Mg II than for Fe II (FWHM of 2700 and 1800 km s$^{-1}$, respectively) that might suggest Fe II origin at somewhat larger distances.

The kinematic width of the line is much narrower than that determined by Forster et al. (2001). They obtained  FWHM = $4500 \pm 1100$ km s$^{-1}$, while we obtained FWHM = $2690 \pm 96$ km s$^{-1}$ in the first observation from the best model fit. All our values of FWHM are much lower than the old measurement. The Forster et al. (2001) measurement was based on the  spectroscopic data with a S/N ratio of 10, and a spectral resolution of 6 - 10 \AA~ while our data have S/N ratio of 50 and a bin size of 0.8 \AA~ in the Mg II range. All our fits also required the change in the value of redshift of the source, in the best fit case from $z = 0.946$ to $z = 0.956$.

The shape of the Fe II pseudo-continuum is best modeled by one of the theoretical templates of Bruhweiler \& Verner (2008). The same template was found to be the best for I Zw I by these authors. This, and the Lorentzian shape of Mg II indicate a considerable similarity between LBQS 2113-4538 and I Zw I, apart from much broader line kinematic width in the first case. It supports the deep meaning in the division of the quasars into A and B class sources, as proposed by Sulentic et al. (2007).

\subsection{Nature of the BLR in Type A quasars}
The Mg II line, despite the very high quality of the data, is well modeled with a single Lorentian shape. We took into account the doublet character of the line 
(this improved the $\chi^2$ by 23.4, although the doublet is not visibly resolved because of the broadening). The quasar is thus similar to NLS1 I Zw I, but the Mg II
line is broader, with FWHM of 2690 km s$^{-1}$. Therefore, the source  probably does not belong to narrow line Seyfert 1  (NLS1) galaxies as defined by Osterbrook \& Pogge (1985). It would be possible to affirm this precisely  with high-quality NIR observations, i.e., with direct measurement of the H$\beta$ properties. In general, H$\beta$ is frequently  somewhat broader than Mg II (e.g., McLure \& Jarvis 2002; Wang et al. 2009). Marziani et al. (2013) showed that the FWHM(Mg II) to FWHM(H$\beta$) ratio may strongly depend on the Eddington ratio, so the expected FWHM of H$\beta$
could thus range from $\sim$ 2100 km s$^{-1}$ (i.e. close to the NLS1 
regime) to $\sim 3200$ km s$^{-1}$. Future observations in the NIR are clearly needed to resolve this question. In any case, the source is a very good representative
of type A quasars, as defined by Sulentic et al. (2007), which likely forms the same physical source class as NLS1, but for objects with much higher black hole mass.

This class of sources is characterized by relatively narrow permitted lines (below 2000 km s $^{-1}$ for NLS1, below 4000 km s$^{-1}$ for bright type A quasars); 
the permitted lines have Lorentzian profiles, strong Fe II emission, relatively small optical variability, but with occasional flares, relatively large X-ray variability, 
and their X-ray spectra are dominated by strong soft X-ray excess, followed by a steep hard X-ray power law (Boller, Brandt \& Fink 1996;
Brandt, Mathur \& Elvis 1997; Leighly 1999; Wang \& Netzer 2003;
Gallo 2006; Panessa et al. 2011; Marziani et al. 2013; Ai et al. 2013). These properties are probably connected with the accretion at the levels close to the Eddington ratio 
(e.g., Pounds et al. 1995; Mineshige et al. 2000; Grupe 2004; Collin et al. 2006).
NLS1 galaxies have bolometric luminosities similar to BLS1 galaxies, but masses on the order of $10^7 M_{\odot}$ (Wandel, Peterson \& Malkan 1999; Grupe \& Mathur 2004; 
Whalen et al. 2006), while type A quasars have luminosities typical for quasars, but masses at the lower end of quasar black hole masses. 
The object studied in this paper also has a black hole mass
in the range $ 0.5 - 1 $ billion solar masses, and the Eddington ratio in the range $ 0.6 - 1.2$, depending on the adopted formula. We do not have any constraints on the 
X-ray properties of this source.

The detailed analysis of the H$\beta$ profiles for NLS1 galaxies and type A quasars strongly suggests a single component, if Lorentzian instead of Gaussians are used 
(e.g., Veron-Cetty et al. 2001, Sulentic et al. 2002, Zamfir at el. 2010). In the FOS/HST I Zw 1 
spectrum of Mg II (which belongs to low ionization lines, like H$\beta$), the Mg II doublet is actually resolved, but there seems to be no need for two separate NLR and BLR
components (Laor et al. 1997). The same conclusion was reached by Shapovalova et al. (2012) who analyzed  H$\beta$ profiles in an 11-year observational campaign of the NLS1 
galaxy Ark 564: the rms and mean profile for  H$\beta$ are identical (but for other lines they are not). However, as  Shapovalova et al. (2012) stress, the decomposition of H$\beta$ into
putative NLR and BLR components is difficult because of the narrowness of the second one.

In LBQS 2113-4538 the Mg II line is broader, so the contribution from the NLR should be easily visible in our high-quality data. When fitting the line profiles, we did not see 
immediate need of an additional narrow feature. We therefore made tests, by fixing the FWHM of the NLR component at 600  km s$^{-1}$ and allowing the normalization to vary. 
In the case of observation 1 we obtained the upper limit of 3 \% at the two-sigma level for this narrow component in Mg II. This  upper limit is not consistent with the usual expectations of
the $\sim 20 $\% contribution of the NLR to permitted lines (Contini et al. 2003, Shapovalova et al. 2012).

It thus seems that in NLS1/Type A quasars not only does the BLR move outward from the black hole as a result of a  high Eddington ratio, but also the NLR moves inward. It does not 
disappear, since the forbidden lines, like [O III] 5007 \AA~ are usually intense in NLS1 objects. This means that the usual gap between BLR and NLR in permitted lines disappears.
This gap exists because of the presence of dust, as argued by Netzer \& Laor (1993). It is possible that the significant change in the broad band continuum shape when passing from
BLS1/Type B sources to NLS1/Type A sources affects the computations presented in that paper.

While a single-component fit for the permitted lines is very interesting, its Lorentzian shape may seem puzzling. This shape is natural in the case of atomic processes, but in AGN
the line shape is determined by the combination of the density and velocity field. A Lorentzian profile, with its broad, extended wings suggests an extended
complex emitter, which leads the $\lambda^{-2}$ shallow decay in the wings thanks to the combination of the two factors.

\subsection{Selection of objects for cosmological application}
The understanding of the formation of BLR in AGN, and in particular of the properties of the Mg II line is important in a much broader context.
As proposed by Watson et al. (2011), BLR size scaling with absolute luminosity offers a possibility to use AGN as probes of the expansion rate
of the Universe. The scaling for H$\beta$ is well studied observationally (see Bentz et al. 2013 for the most recent studies), and the same scaling is expected for Mg II on the basis of the theoretical understanding of the scaling mechanism (Czerny \& Hryniewicz 2011).  This line can be used for optical monitoring of medium redshift 
quasars but attention should be paid to the proper selection of the sources based on the line shape in order to avoid unwanted line variability unrelated to the 
changes in the continuum (see Czerny et al. 2013). The quasar LBQS 2113-4538 belongs to the sources which show clear Mg II variability, like NGC 3783 (Reichert et al. 1994), and NGC 4151 
(Metzroth et al. 2006), while other objects monitored in Mg II did not show the expected variability, like PG 1247+268 (Trevese et al. 2007). On average, class A sources are less variable in the optical band (e.g., Papadakis et al. 2000; Klimek et al. 2004, Ai et al. 2013)
but the example of LBQS 2113-4538 shows that this is not always the case, and the simplicity of the Mg II shape in these sources simplifies the decomposition of the spectrum and the 
overall monitoring. The Eddington ratio in the source is close to 1, nevertheless, we do not see  evidence of the Mg II line coming from wind outflow, which would 
prevent the use of the line for black hole mass and intrinsic luminosity measurement.


\begin{acknowledgements}
We thank the anonymous referee for valuable comments, and in particular for calling our attention to the importance of the correct doublet ratio value for the determination of the Fe II template. Part of this work was supported by the Polish grant Nr. 719/N-SALT/2010/0, and the spectroscopic 
observations reported in this paper were obtained with the Southern African Large Telescope (SALT),  
proposal SALT/2012-1-POL-008. KH, BC, MK and AS
acknowledge the support by the Foundation for Polish Science through the
Master/Mistrz program 3/2012. The OGLE project has received funding from the European Research Council
under the European Community's Seventh Framework Programme
(FP7/2007-2013) / ERC grant agreement no. 246678 to AU.
The Fe II theoretical templates described in 
Bruhweiler \& Verner (2008) were downloaded from the
web page http://iacs.cua.edu/personnel/personal-verner-feii.cfm with the permission of the authors. 
This research has made use of the NASA/IPAC Extragalactic Database (NED) which is operated by the 
Jet Propulsion Laboratory, California Institute of Technology, under contract with the National Aeronautics and Space Administration. 
\end{acknowledgements}

\end{document}